# Increase of SRG modulation depth in azopolymers-nanoparticles hybrid materials

R. Falcione[a], M. V. Roldan[b], N. Pellegri[b], S. Goyanes[c], S. Ledesma*[a], M.G. Capeluto[a]

a *Laboratorio de Optica y Fotónica (LOFT), Departamento de Física, Facultad de Ciencias Exactas y Naturales, Universidad de Buenos Aires and IFIBA, CONICET, Cuidad Universitaria, Buenos Aires, 1428, Argentina* b *Laboratorio de Materiales Cerámicos, Instituto de Física Rosario, Universidad Nacional de Rosario, CONICET, Rosario, Argentina.*

c *Laboratorio de Polímeros y Materiales Compuestos (LPM&C), Departamento de Física, Facultad de Ciencias Exactas y Naturales, Universidad de Buenos Aires and IFIBA, CONICET, Cuidad Universitaria, Buenos Aires, 1428, Argentina*

*email: ledesma@df.uba.ar

**Abstract:** Thin films of azopolymer-nanoparticles hybrid materials were fabricated with poly[1-[4-(3-carboxy-4-hydroxyphenylazo) benzenesulfonamido]-1,2-ethanediyl] (PAZO) and different concentrations of Ag and AgAu nanoparticles (NPs). By illuminating the films with polarized interference patterns, surface relief gratings (SRGs) were recorded. It was found that for some concentrations of NPs their modulations and diffraction efficiency were higher than the obtained for PAZO films without NPs. The effect was mainly explained by the increase of the free volume available for the photoisomerization for certain concentrations of NPs. The dependence of the diffraction efficiency on concentration was directly related to changes in modulation depth. When doping with NPs, the maximum efficiency increases more than two times the efficiency without NPs.

1. Introduction

Inorganic-in-organic hybrid material have been extensively studied in the past years because it extends the range of possible applications [1]. Different types of materials can be combined to generate a new material that not only inherits the properties of both compounds but also could develop new properties according to the chemical or physical interactions between them. Azobenzene containing polymers (azopolymers) are organic polymers with a photosensitive response, under an appropriate illumination they produce macroscopic and microscopic changes. Hybrid materials can be generated not only to improve their optically induced properties, conductivity or hardness but also to produce new functionalities.

The possibility to induce its physical properties by illumination makes azopolymers very useful for applications such as optical memory storage [2], lithography [3, 4], optical switching [5] and actuators of nanoscale objects [6, 7]. It is well known that the photoisomerization of the azo compound is the phenomenon that triggers the changes in optical properties. When the azobenzene photoisomerizes, a rotation of the azo group occurs producing a conformational change between the two isomers cis and trans. For a polarized incident beam the azobenzene will perform several random rotations until its transition dipole moment is oriented perpendicular to the electric field of the incident beam, where the probability of absorbing photons is zero. Macroscopically speaking, when the azobenzenes are included in the polymer matrix, this reorientation process, known as photoselective orientation (or angular hole burning), produces an optical anisotropic material leading to birefringence ($\Delta n$) and optical dichroism. The rate at which all the isomers reach this state of orientation depends on the beam intensity and the temperature, among other intrinsic parameters such as quantum efficiency of photoisomerization and loss rate due to thermally activated angular diffusion [8].

Another interesting phenomenon that is observed in some azopolymers and still needs further understanding is mass migration. It occurs when the azopolymer is illuminated by a structured light pattern, with intensity, polarization or phase spatial variations. By illuminating the azopolymer with spatial variating beam, complex structures composed not only by mass relief, but also of ordered/disordered molecules patterns that give rise to adhesion and a charge spatial distribution, build up in the surface of the azopolymer [9, 10]. Notably this mass migration occurs at room temperature (RT) where the polymer behaves as a glass (RT is tens of degrees lower than the glass transition temperature $T_G$, for example $T_G \sim 95$ °C for PAZO [11]). Normally mass migrates from bright illuminated areas to dark ones with a velocity that is proportional to the gradient of the intensity [12, 13]. This phenomenon is typically studied by illuminating the material with an interference pattern to induce a surface relief grating (SRG) and simultaneously measuring the diffraction efficiency with another laser at a frequency that doesn't produce interactions with the azopolymer.

In order for the azobenzene to perform the fast movements that occur during photoisomerization and the reorientation of the transition dipole moment, a certain free volume in the polymer matrix should be available [8]. Free volume can be created as static voids by inefficient chain packing or transient voids by thermal diffusion of the polymer segments. The larger the free volume, the easier for molecules to diffuse through the polymer. Merkel showed that nanoparticles (NPs) with diameter smaller than 50nm can act as nanospacers to prevent the polymer chain from packing close [14], then increasing the number and size of the free volume elements, and consequently allowing higher and faster transport through the polymer. Regarding the azobenzenes in the polymer matrix, increasing the size and number of the free volume elements would allow higher and faster diffusive rotation, reorientation and photoisomerization, making both the photoisomerization rate and the number of allowed azobenzene available to isomerize higher.

Metallic NPs can be used as spacers, they are easy to get in with different radii in a monodisperse solution. Remarkably, they are also ideal to generate hybrid materials with novel optical properties since their optical response is tunable by size, shape, and the complex refractive index of the NP and the medium that surrounds it. In fact, light interaction with NPs at the localized surface plasmon polariton resonance (LSPPR) produces high electromagnetic fields around the NP, an effect directly related to the enhancement of the Raman signal. At this frequencies, NPs absorption of light is very strong, leading to heat due to Joule effect.

Regarding azopolymer-metallic NPs hybrids, some results previously published showed increased values of the photo-orientation rate ($\tau$) and the maximum birefringence ($\Delta n_{max}$), for certain concentrations of Ag and Au NPs [15-17]. In general, the results were explained phenomenologically by the effect of LSPPR and the scattering in the NPs. On the one hand, at the plasmon resonances the field enhancement and the temperature increase in the vicinity of the NPs, would favor the azobenzene mobility and enhance the photoisomerization rate. On the other hand, the scattering of light by the NPs would allow to excite and reorient the originally off-plane chromophores that in other case, would not contribute to the birefringence [18]. Zhou et al. observed a maximum in $\tau$ and in $\Delta n_{max}$ for an amount of Ag NPs of 10 ng [16]. Measurements performed at different wavelengths allowed to infer phenomenologically that LSPPR in Ag NPs immersed in the polymer matrix would be the key factor for the improvement of the optical properties. However, Shen et al. observed a different tendency on the photo-orientation rate and $\Delta n_{max}$ for different azopolymers [17]. In the case of AzoCN these magnitudes had a minimum due to the cross-linking effect of the azopolymer around the NPs that restrain the mobility of the azobenzene molecules and reduce accordingly the photo-orientation rate. This effect wasn't observed in AzoCH3, where $\tau$ and $\Delta n_{max}$ monotonically increased. On the other hand, recently Sekkat et al. measure an increased birefringence and an absorbance enhancement in azobenzene – Au NPs [19].

Hybrids of azopolymers and nonmetallic NPs also showed improvements in the photoinduced birefringence. The increase of the birefringence in hybrids of $TiO_2$ NPs and PAZO with different thermal treatments was explained by reorientation of the off-plane chromophores [20]. The measurements exhibited a maximum in $\Delta n_{max}$ for some concentrations. The annealed samples showed higher value of $\Delta n_{max}$, but non-annealed samples showed lower response time $\tau$. Similar results were observed in the azopolymer $P_1$ doped with different concentrations of $SiO_2$ NPs [18]. It was suggested that NPs might reduce the interaction between the azobenzene molecules close to the surface of the NP resulting in enhanced mobility and higher birefringence. This is quite logical reasoning since as it was said before, azobenzene requires certain free volume to isomerize. Moreover, the increase of the free volume using particles acting as spacers of polymer chains would be a process dependent on the size of the NPs, but not on their chemical composition or refractive index. On the other hand, the scattering of light by the NPs, is an effect that depends on size and composition. As it was stated by Nazarova et al. [18], scattering would allow to excite and reorient the originally off-plane azochromophores that would not contribute to $\Delta n$. If that is the case an increase on $\Delta n$ of doped samples could be observed. $P_1$ and $P_{1-2}$ films doped with ZnO NPs showed higher (lower) $\Delta n_{max}$ for smaller (larger) NPs. It is suggested that smaller NPs allow more free volume, leading to higher $\Delta n_{max}$, while large NPs increase scattering which reduces the transmitted intensity, and therefore the effective value of $\Delta n_{max}$ [21].

Despite the previous work, these arguments have been only used to phenomenologically explain changes in absorption and birefringence, where the measured enhancement has great potential for optical memory storage or optical switching. A less explored effect, or even unexplored is the mass migration in hybrid materials. Berberova et al. [22] showed by recording holographic polarization gratings that both diffraction efficiency and modulation of the SRG increased by adding ZnO NPs to PAZO at several concentrations, maybe caused by an enhanced mobility of the azos in the vicinity of the NPs. Mass transfer of $SiO_2$ NPs induced by holographic recording in an azopolymer was observed by Y. Tomita et al [23].

In this work we study the mass transport in hybrids PAZO and two different compositions of nanoparticles: Ag and AgAu (30% Ag /70% Au). We characterized the diffraction efficiency of the recorded SRGs on films fabricated with different concentrations of NPs. For both types of NPs, we observed that diffraction efficiency can be maximized for certain concentrations. By proposing a simple model for the efficiency as a function of the induced modulation depth of the SRG, we observed that the increase of the efficiency is directly related to the increase of the modulation depth. The results are adequately explained in terms of the increased free volume for low concentrations and scattering for high concentrations.

## 2. Experimental

### 2.1 Materials

A commercial azopolymer poly[1-[4-(3-carboxy-4-hydroxyphenylazo) benzenesulfonamido] -1,2-ethanediyl] (PAZO) from Sigma Aldrich was used. The solvent (methanol, reagent grade) and the plasticizer (ethylene glycol) were purchased from Biopak (Argentina). The following reagents were used for the nanoparticles synthesis: AgNO3 from Merck, HAuCl4·3H2O and [N-[3-(trimethoxysilil)propyl] diethylenetriamine] from Sigma-Aldrich and absolute ethanol from Cicarelli.

Two types of NPs were specially designed and fabricated for this experiment with the aim of obtaining different overlaps between the NPs and PAZO absorption spectra. Ag NPs were obtained by chemical synthesis. Briefly, 68 mg of $AgNO_3$ were dissolved in 75 mL of absolute ethanol by ultrasound stirring. Separately, 0.644 mL of [N-[3-(trimethoxysilil)propyl] diethylenetriamine] were mixed by magnetic stirring with 25 mL of absolute ethanol under N2 atmosphere. Then, both solutions were mixed in a single neck round bottom flask equipped with a vertical condenser, N2 atmosphere and magnetic stirring and it was putted in a temperature bath at 60 ºC during 5 h. After this time, a homogeneous yellow solution was obtained. The obtained Ag concentration was equal to 4 mM and it was adjusted by dilution or solvent evaporation according to the use requirement. Colloidal AgAu NPs were obtained as follows: Firstly, a 25 mM $Au^{3+}$ stock solution was obtained by dissolving 194 mg of $HAuCl_4·3H_2O$ in 20 mL of absolute ethanol. Then, the AgAu NPs synthesis started with the dissolution of 644 μL of [N-[3-(trimethoxysilil)propyl]diethylenetriamine] in 17.8 mL of absolute ethanol with magnetic stirring under $N_2$ atmosphere. On the other hand, 0.0191 g of $AgNO_3$ were dissolved in 80 mL of absolute ethanol and following 1.83 mL of $Au^{3+}$ stock solution was added. Then, both solutions were mixed in a single neck round bottom flask equipped with a vertical condenser, N2 atmosphere and magnetic stirring and it was putted in a temperature bath at 60 ºC during 5 h. After this time, a homogeneous reddish suspension was obtained corresponding to colloidal AgAu nanoparticles. The final concentration of total metal (Ag + Au) was 1.57 mM and it was adjusted by adding or evaporating solvent as appropriate. Fig. 1 shows the UV-Vis absorption spectra for Ag and AgAu (70/30), both with 5 nm diameter, at equal molar concentration. The absorption maxima are centered in 408 nm (Ag) and 458 nm (AgAu). At 473 nm (the excitation laser wavelength) Ag NPs have higher absorption than AgAu NPs.

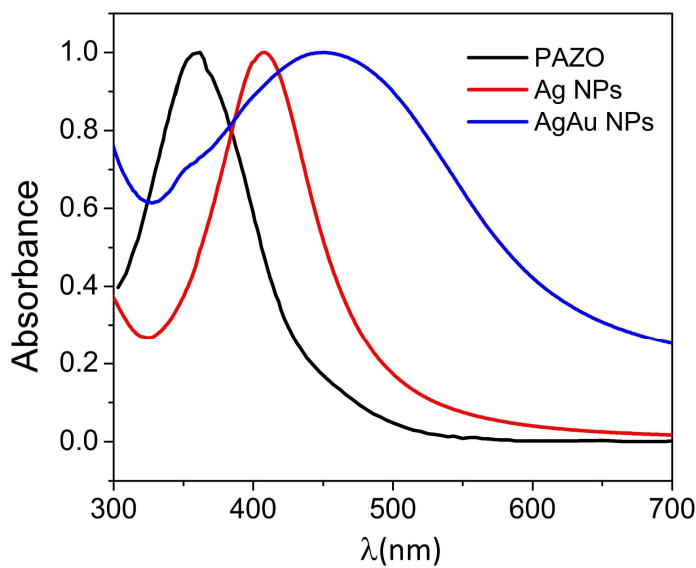

Fig. 1. Normalized UV-Vis absorption spectra of Ag NPs (blue), AgAu NPs (red) and PAZO (black).

**2.2 Film preparation and characterization**

The photoinduced properties of PAZO/NPs for different concentrations of NPs were studied by considering two sets of samples: one with Ag NPs and another with AgAu NPs. For each concentration and type of NPs, the films were prepared by dissolving 105 mg of PAZO in 1 ml of methanol and 0.3 ml of ethylene glycol. Then, 200 µl of the NPs suspension were added to the solution. After being mixed 30 min in an ultrasonic bath, the films were spin coated onto a clean glass coverslip. Droplets of 100 µl of PAZO/NPs solution were deposited on the coverslip and let it spin for 1 min, at a speed of 3600 rpm. This process was done three times in order to get thick films. The solvent was evaporated by heating the sample in a furnace. A stepped rising temperature protocol was used to produce homogeneous films without bubbles. The sequence of temperatures was 24 h at 50 °C, 2 h at 60 °C and 2 h at 80 °C. A third set of samples with the described amounts of PAZO, ethylene glycol and methanol plus 200 µl of ethanol was prepared to perform a control experiment. The concentrations of metals in the NPs suspensions were 0, 1.5, 3, 4.5, 6 mM, for both types of NPs. The mass concentration of metals with respect to that of PAZO is 0, 0.03, 0.06, 0.09, 0.12 %, for samples with Ag NPs, and 0, 0.04, 0.08, 0.12, 0.15 % for samples with AgAu NPs. The thicknesses of the resulting films were measured from cryogenic fractures of the sample using Scanning Electron Microscopy (SEM) performed by a Carl Zeiss NTS-Supra 40, for which we sputtered a 10 nm layer of platinum to avoid the sample to charge. The average thickness was measured to be about 700 nm. The UV-vis absorption spectrum of a typical sample without NPs is shown in Fig. 1, it consists of a wide band peak centered in 360 nm with a FWHM of 100 nm. As it can be seen, the absorption spectra are not shifted enough to observe a distinctive plasmonic effect for different types of nanoparticles. Further studies are being carried out in this direction.

### 2.3 SRG recording.

In order to study mass transport a Lloyd's Mirror interferometer was used, as it is schematized in Fig 2. In this interferometer two halves of the beam interfere to produce a sinusoidal interference pattern on the sample. In this wavefront division interferometer half of the beam impinges directly on the sample and the other reflects in the mirror before impinging on the sample. The fringe period $\Lambda = \lambda_b/2 \sin(\theta)$ depends on the incidence angle ($\theta$) and the wavelength of the recording light ($\lambda_b$ = 473nm).

A diode laser was filtered and expanded using a 10x objective Melles Griot (O), a lens (L), and a 50 µm diameter pinhole (PH), resulting in a beam with a FWHM of 2.9 mm. A linear polarizer ($P_1$) was used to set the polarization of the interference pattern, and its transmission axis was set along the $\hat{x}$ direction. A half wave plate (HWP) was used to maximize the intensity over the sample. The mirror (M) was set at an angle $\theta \sim 5°$ with respect to the incident beam direction, resulting in a fringe period of about 2.7 µm. The sample (S) was aligned perpendicular to the mirror. The average power density over the sample is 0.2 W/cm². $\vec{u}$ is the coordinate along the sample.

In this polarization configuration and for an incidence angle smaller than π/4, the resulting polarization on the sample is linear and perpendicular to the fringes, and the light intensity has a large modulation. The fringes contrast ($C$) results

$$C = \frac{I_{max}-I_{min}}{I_{max}+I_{min}} = \frac{2\rho}{1+\rho^2} \cos(2\theta), \qquad (1)$$

where $I_{max}$ and $I_{min}$ are the maximum and minimum values of the incident intensity, and $\rho$ is the mirror reflectance at the incidence angle $\theta$ for photons of wavelength $\lambda_b$. In this experiment we used a protected aluminum mirror for which the reflectivity is $\rho = 0.92$. From Eq. (1) it is seen that the contrast is higher for smaller angles.

The SRG recording was monitored in real time by measuring the intensity of the diffraction orders 0 and 1 ($I_0$ and $I_1$, respectively) of a He-Ne ($\lambda_p$= 632.8 nm) laser beam. The polarization of the probe was set at 45° from the $\hat{y}$ direction using a second linear polarizer (P₂). The intensity of the diffraction orders was measured with two photodiodes (D₀, D₁). The first order diffraction efficiency was computed as the quotient between $I_1$ and the input intensity.

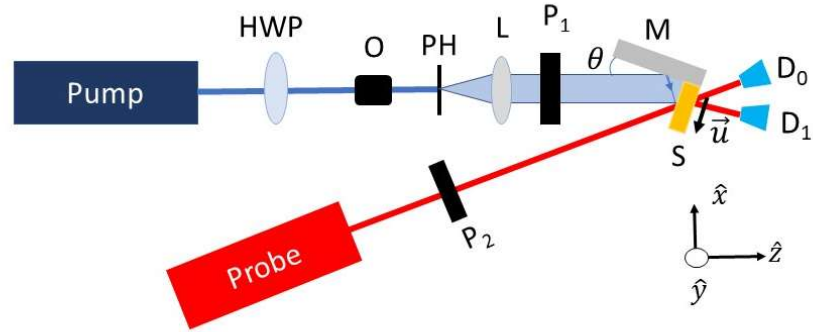

Fig. 2. Experimental set up for recording the SRGs. A pump laser with 473 nm, a half wave plate (HWP), an objective (O), a pinhole (PH), a lens (L), a linear polarizer with its transmission axis set along the $\hat{x}$ direction ($P_1$), a mirror (M) set at an angle $\theta$ with respect to the pump beam, the sample (S), a probe laser (632.8 nm), a linear polarizer ($P_2$) oriented at 45° from the $\hat{y}$ direction, photodiodes ($D_0$, $D_1$). $\vec{u}$ is the coordinate along the sample.

### 3. Results and discussion

The SRGs were recorded on a relatively large area of the sample of about 14 mm². The films were illuminated with an interference pattern for 10 h. The temporal evolution of the diffraction efficiency of a typical sample is shown in Fig. 3. The efficiency grows continuously without reaching a saturation value during the time the measurement lasts.

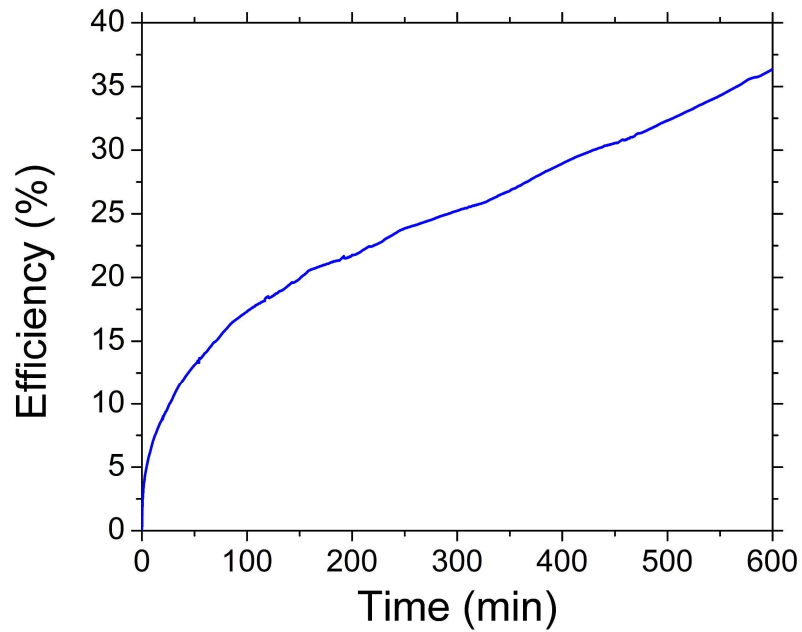

Fig. 3. Temporal evolution of the diffraction efficiency for a typical sample.

The SRG modulations were measured from the cryogenic cuts of the samples using an SEM. Fig. 4 shows typical images, where Fig. 4(a) is the film before recording the SRG, and

Fig. 4(b) shows an SRG with low modulation. The recorded SRG consistently agrees with the illumination pattern, as it consist of a sinusoidal mass modulation with a period of about 2.7 um as it is expected from the experimental configuration.

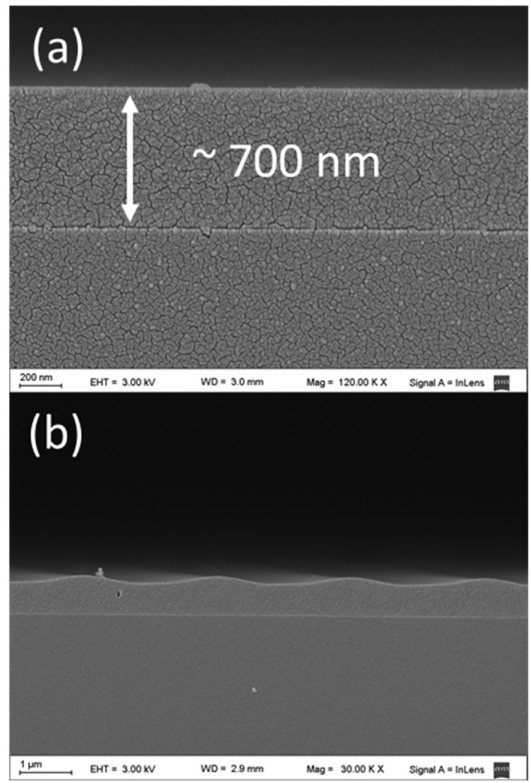

Fig. 4. SEM images of the cryogenic cut of a typical sample: (a) before recording the SRG, and (b) a low modulation SRG.

Films with all concentrations of NPs were illuminated and the intensity of the first order of diffraction ($I_1$) was measured in real time. After 10 h of exposition to 473 nm interference pattern, SRGs modulations were measured from SEM images. The modulation average for each set of samples at different concentrations are shown in Fig. 5. With the aim of comparing, the modulation average for samples without NPs was also plotted. The error bar for each concentration corresponds to dispersion of the measured values for the modulation, for the set of samples of such concentration. Considering these error bars, for low concentrations of NPs, the modulations for samples with NPs are similar to those for samples with no NPs. For both types of NPs, there is a maximum of modulation and its value is larger than for samples without NPs. The concentrations at which such maxima occur are about 0.08 % for AgAu NPs, and 0.09 % for Ag NPs. For higher concentrations of NPs, the modulation drops for both types of NPs. Despite the fact that the error bars are large, the maximum for samples with NPs is clearly distinguishable from the modulation of the samples without NPs. This can be interpreted as follows. As it is well known, the SRG is formed due to the mass transport induced by the photoisomerization of the azo groups. A necessary condition for photoisomerization to occur is the availability of free volume. The incorporation of the NPs to the azopolymer increases the free volume: they reduce the interaction between the azo groups in their vicinity separating the polymer chains, which increases the free volume [14]. Considering that the smaller distance between two polymer chains is about the length of the azo-chromophore (~1nm) [24], NP with 5 nm diameter would increase the separation between the two chains. Then it is reasonable to think that the higher the concentration, the higher the increase in the free volume. As the free volume increases there are more azo groups available to isomerize, favoring mass transport. Eventually, for much larger concentrations of NPs the opposite effect occurs: NPs occupy larger space reducing the number of azo groups per unit of volume. This possible influence of the NPs in the available free volume is in accordance with the observations of Nazarova [18, 21, 15] about the changes in the induced birefringence of samples doped with NPs.

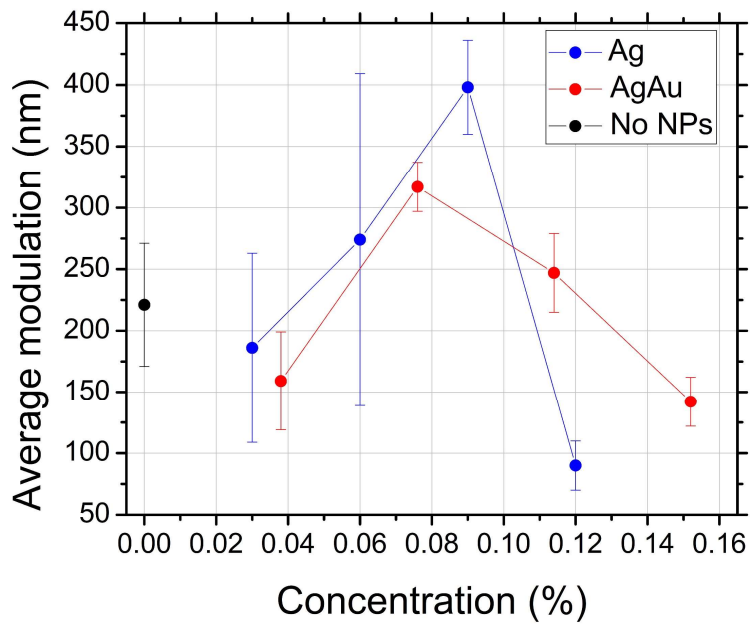

Fig. 5. SRGs modulation vs. NPs concentration. Samples with different types of NPs are shown: Ag (blue), AgAu (red), no NPs (black).

In addition, the first order diffraction efficiency and its dependence on the SRG modulation is analyzed. In Fig. 6 the average values for efficiencies vs the modulation at each concentration are shown. As it is expected, for a phase grating the efficiency increases as the modulation grows for phase differences smaller than 1.84, considering that the efficiency of the diffracted orders is ruled predominantly by Bessel functions. As it is well known, azopolymer SRG gratings consists of a phase grating constituted by a mass grating with variable height and 90º phase shifted birefringence grating.

In order to model this effect, and in accordance with Hwang et al. [25] suppose that two gratings are being developed. On the one hand, the height of the surface grating can be modeled as

$$y(u,t) = d + \frac{m}{2}\sin(\Lambda\, u), \qquad (2)$$

where $d$ is the film thickness, $m$ is the grating modulation, and $u$ is the coordinate on the sample perpendicular to the direction of the interference fringes. On the other hand, the refractive index $n(u,t)$ also varies along the $\vec{u}$ direction due to the induced birefringence. In order to model the refractive index, the material refractive index $n_0$ and the time dependent photoinduced birefringence $\Delta n(t)$ are considered. A sinusoidal spatial variation with the period of the fringes $\Lambda$ is considered. Consequently, the proposed expression for the refractive index is

$n(u,t) = n_0 + \frac{1}{2}\Delta n(t)(1 + \sin(\Lambda u + \pi))$. The transmittance function of the recorded material is

$$T(u,t) = e^{ik(n(u,t)-1)y(u,t)}, \qquad (3)$$

with $k = \frac{2\pi}{\lambda_p}$. For this transmittance, a Fourier propagation will give a first order diffraction efficiency of

$$E_1(u,t) = \left| J_0\left(\frac{m}{8} k\, \Delta n(t)\right) J_1\left(k\left(\frac{m}{2}(n_0 - 1) + \frac{1}{2}\left(\frac{m}{2} - d\right)\Delta n(t)\right)\right)\right|^2. \qquad (4)$$

There are some experimental considerations that must be done in order to fit the model to the data. On the one hand, since birefringence saturates in time scales much shorter than the growth rates for the mass grating [9], the saturation value for the birefringence is considered instead of its time dependence $\Delta n(t)$. On the other hand, the thickness of the film is considered as an effective thickness. Taking a look at Eq. (4), the thickness $d$ only appears accompanied by the birefringence, meaning that it is related to the thickness of the birefringence grating. However, mass transport has a disorder effect in the molecules near the film surface where the azopolymer molecules translates, while the ones closer to the substrate are much less affected by such transport. Thus, it is expected that the birefringence grating is formed in a smaller thickness than the thickness of the initial film. Also, even though it is attempted to characterize the SRG in the regions were the modulation is higher (near the edge of the mirror and in the middle of the sample), the SEM localization accuracy of small areas in big samples is poor. Then the thickness $d$ and the modulation $m$ are varied around their nominal values. Then these parameters are optimized considering the whole set of data. Last, the refractive index is considered to be 1.75 as it was measured in Berberova et al. [26]. The fitting parameters turned out to be $\Delta n = 0.09$, and $d$ = 500 nm. This simple model shows good accordance with the data and allows to conclude that the only effect that conducts to the increase of the efficiency is the SRG modulation growth.

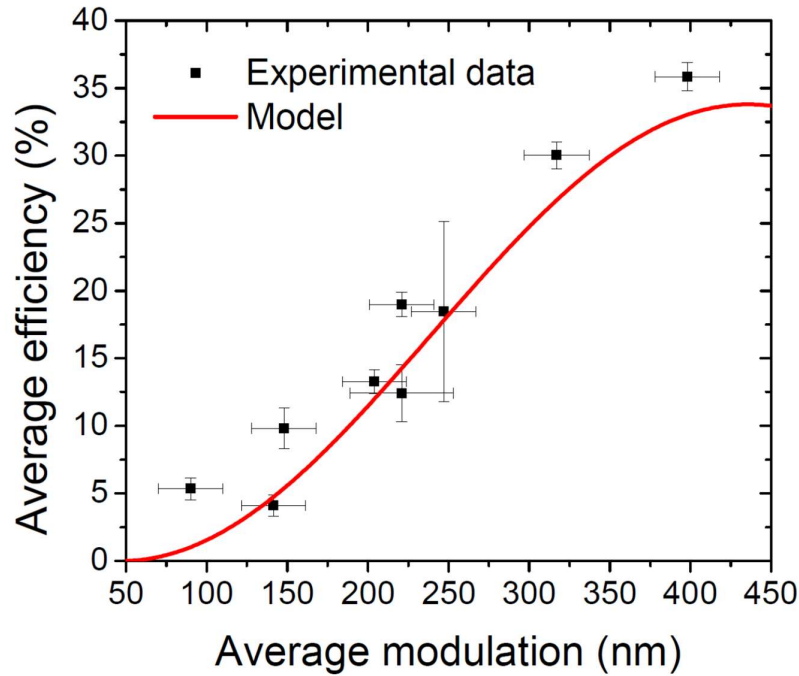

Fig. 6. Average efficiency of the first order of diffraction vs. the SRG average modulation (experimental data, blue dots), and theoretical curve (red line).

**Conclusions**

The effect of the concentration and type of NPs on the photoinduced properties of azopolymers-nanoparticles hybrid materials was studied. Four different concentrations of Ag and AgAu NPs were combined with PAZO. SRGs were recorded by laser illumination on the hybrid materials, and the first order diffraction efficiency was simultaneously measured. Efficiency grew continuously while the sample was illuminated, and the value reached after 10 h of illumination showed a dependence on the concentration. The modulation of the photoinduced SRGs were measured from the cryogenic fractures of the samples using an SEM. It was found that the SRGs recorded in films of PAZO/NPs reached higher modulations than the SRGs recorded in PAZO films. For both types of NPs there is a concentration for which the modulations are maximal. A possible explanation for this growth is that the NPs would acts as spacers of the polymer chains increasing the number and size of the free volume elements, allowing more azo groups to be available for isomerization, and therefore favoring the mass transport. By modeling the diffraction efficiency dependence on the modulation, it was observed that the origin of the behavior of the efficiency was directly related to the modulation. The model considered a phase grating produced by the SRG and the variations on the birefringence. This suggests that the only cause of the growth of such diffraction efficiency is the increase of the SRG modulation. These results prove the possibility of improving the optical response of the material by means of the addition of a controlled concentration of metallic nanoparticles. This improvement is directly related to an increase on the modulation depth.

**Acknowledgments**

The authors would like to thank the financial support from the University of Buenos Aires (UBACyT 20020170100381BA and 20020170100564BA), ANPCyT (PICT-2018-04235, PICT 2017-2362), and CONICET (PUE-IFIR-RD 1691/16 and 11220150100475CO)